\documentclass[onecolumn,showpacs,preprintnumbers,amsmath,amssymb,eqsecnum]{revtex4}
\usepackage{dcolumn}
\usepackage{bm}
\usepackage[dvips]{graphicx}
\usepackage{wrapfig}
\usepackage{psfrag}
\begin{document}
\Large

\def\sh{\mathop{\rm sh}\nolimits}
\def\ch{\mathop{\rm ch}\nolimits}
\def\var{\mathop{\rm var}}\def\exp{\mathop{\rm exp}\nolimits}
\def\Re{\mathop{\rm Re}\nolimits}
\def\Sp{\mathop{\rm Sp}\nolimits}
\def\kp{\mathop{\text{\ae}}\nolimits}
\def\bk{{\bf {k}}}
\def\bp{{\bf {p}}}
\def\bq{{\bf {q}}}
\def\lra{\mathop{\longrightarrow}}
\def\Const{\mathop{\rm Const}\nolimits}
\def\sh{\mathop{\rm sh}\nolimits}
\def\ch{\mathop{\rm ch}\nolimits}
\def\var{\mathop{\rm var}}

\def\Re{\mbox {Re}}
\newcommand{\Z}{\mathbb{Z}}
\newcommand{\R}{\mathbb{R}}
\def\mK{\mathop{{\mathfrak {K}}}\nolimits}
\def\mR{\mathop{{\mathfrak {R}}}\nolimits}
\def\mv{\mathop{{\mathfrak {v}}}\nolimits}
\def\mV{\mathop{{\mathfrak {V}}}\nolimits}
\def\mD{\mathop{{\mathfrak {D}}}\nolimits}
\def\mN{\mathop{{\mathfrak {N}}}\nolimits}
\def\mS{\mathop{{\mathfrak {S}}}\nolimits}
\newcommand{\ccm}{{\cal M}}
\newcommand{\cE}{{\cal E}}
\newcommand{\cV}{{\cal V}}
\newcommand{\cI}{{\cal I}}
\newcommand{\cR}{{\cal R}}
\newcommand{\cK}{{\cal K}}
\newcommand{\cH}{{\cal H}}

\def\br{\mathop{{\bf {r}}}\nolimits}
\def\bS{\mathop{{\bf {S}}}\nolimits}
\def\bA{\mathop{{\bf {A}}}\nolimits}
\def\bJ{\mathop{{\bf {J}}}\nolimits}
\def\bn{\mathop{{\bf {n}}}\nolimits}
\def\bg{\mathop{{\bf {g}}}\nolimits}
\def\bv{\mathop{{\bf {v}}}\nolimits}
\def\be{\mathop{{\bf {e}}}\nolimits}
\def\bp{\mathop{{\bf {p}}}\nolimits}
\def\bz{\mathop{{\bf {z}}}\nolimits}
\def\bbf{\mathop{{\bf {f}}}\nolimits}
\def\bb{\mathop{{\bf {b}}}\nolimits}
\def\ba{\mathop{{\bf {a}}}\nolimits}
\def\bx{\mathop{{\bf {x}}}\nolimits}
\def\by{\mathop{{\bf {y}}}\nolimits}
\def\br{\mathop{{\bf {r}}}\nolimits}
\def\bs{\mathop{{\bf {s}}}\nolimits}
\def\bH{\mathop{{\bf {H}}}\nolimits}
\def\bk{\mathop{{\bf {k}}}\nolimits}
\def\be{\mathop{{\bf {e}}}\nolimits}
\def\bnul{\mathop{{\bf {0}}}\nolimits}
\def\bq{{\bf {q}}}

\newcommand{\oV}{\overline{V}}
\newcommand{\vkp}{\varkappa}
\newcommand{\os}{\overline{s}}
\newcommand{\opsi}{\overline{\psi}}
\newcommand{\ov}{\overline{v}}
\newcommand{\oW}{\overline{W}}
\newcommand{\oPhi}{\overline{\Phi}}

\def\mI{\mathop{{\mathfrak {I}}}\nolimits}
\def\mA{\mathop{{\mathfrak {A}}}\nolimits}

\def\st{\mathop{\rm st}\nolimits}
\def\tr{\mathop{\rm tr}\nolimits}
\def\sign{\mathop{\rm sign}\nolimits}
\def\d{\mathop{\rm d}\nolimits}
\def\const{\mathop{\rm const}\nolimits}
\def\O{\mathop{\rm O}\nolimits}
\def\Spin{\mathop{\rm Spin}\nolimits}
\def\exp{\mathop{\rm exp}\nolimits}

\def\mI{\mathop{{\mathfrak {I}}}\nolimits}
\def\mA{\mathop{{\mathfrak {A}}}\nolimits}

\def\st{\mathop{\rm st}\nolimits}
\def\tr{\mathop{\rm tr}\nolimits}
\def\sign{\mathop{\rm sign}\nolimits}
\def\d{\mathop{\rm d}\nolimits}
\def\const{\mathop{\rm const}\nolimits}
\def\O{\mathop{\rm O}\nolimits}
\def\Spin{\mathop{\rm Spin}\nolimits}
\def\exp{\mathop{\rm exp}\nolimits}

\title{One more variant of discrete gravity having "naive" continuum limit}

\author {S.N. Vergeles\vspace*{4mm}\footnote{{e-mail:vergeles@itp.ac.ru}}}

\affiliation{{Landau Institute for Theoretical Physics, Russian
Academy of Sciences,}\linebreak Chernogolovka, Moskow region,
142432 Russia }

\begin{abstract}
Some variant of discrete quantum theory of gravity having "naive"
continuum limit is constructed. It is shown that in a highly
compressed state of universe a sort of "high-temperature
expansion" is valid and, thus, the confinement of "color" takes
place at early stage of universe expansion. In the considered
theory any nontrivial representation of the local Lorentz group
(i.e. spinor, vector and so on fields) play the role of color. The
arguments are given in favor of a significant noncompact packing of
quantized field modes in momentum space.
\end{abstract}


\maketitle

\section{INTRODUCTION}

Since the traditional methods of quantization of gravity in four
dimensions prove to be inconsistent because of ultraviolet
divergences, a natural idea has arisen that qualitatively
different physics takes place at supersmall distances (of the
order of the Planck length and smaller).

At present, a predominant opinion among theoretical physicists is
that superstring theory is a fundamental physical theory. In a
ten-dimensional space, superstring theory is self-consistent.
The superstring theory involves gravitational interaction.

However, one encounters an extremely difficult problem within the
string ideology, the problem of the compactification of six
dimensions and the construction of a long-wavelength physics in
four dimensions. Therefore, recently, actual progress in solving
many problems of the quantum theory of gravity and quantum
cosmology along this line has not been made.

The aforesaid justifies the existence of certain other ideas
underlying the fundamental quantum field theory and, first of all,
the theory of gravity. In our opinion, the most interesting idea
is the idea of discrete space-time, which is the main subject of
the present paper \footnote{We mention here one more attempt of
construction of self-consistent quantum theory of gravity which is
based on ideas elaborated under the study of ${}^3He$ in
superfluid phase: G. E. Volovik, E-print archives gr-qc/0101111;
"The Universe in a
 Helium Droplet", Clarendon Press. Oxford 2003.}.

The idea of the discreteness of space-time (as applied to the
theory of gravity) was first formulated in the pioneering work by
Regge \cite{1} long before the appearance of string theory.
According to Regge, the role of smooth Riemannian spaces is played
by piecewise flat spaces, namely, simplicial complexes (necessary
information from the theory of simplicial complexes is given at
the beginning of the next section). To each one-dimensional simplex
(edge) is assigned its length, so that, if the set of three edges
forms a boundary of a two-dimensional simplex (a triangle), then
the lengths of these edges satisfy the triangle inequalities.
Thus, the geometry of a complex is completely defined. The
quantity representing an analogue of the Riemann tensor on a
smooth Riemannian space proves to be nonzero only on a set of
$(D-2)$-dimensional simplices ($D$ is the dimension of the
simplicial complex); i.e., the curvature tensor becomes a
distributions.

The detailed account of the Regge calculus is given in
\cite{2}. An approach to discrete geometry similar to the
Regge calculus can also be found in \cite{3},\cite{4}.

Despite the obvious elegance of the Regge calculus, this theory
proves to be very inconvenient when passing to quantum
theory. Indeed, the independent variables that determine
Regge action are the lengths of one-dimensional simplices
subject to a large number of constraints, namely, to
triangle inequalities. Moreover, the introduction of Dirac
fields to the theory creates a new difficulty consisting in
the absence  of orthonormal bases in the explicit form.
Possibly, this is the reason why the variant of discrete
gravity based on the so-called $B$-$F$-theory is currently
being developed more intensively.

The $B$-$F$-theory is developed on the basis of the action for
gravity theory in the Palatini form.
I write out the action for a four-dimensional
gravity theory with a massless Dirac field, especially because
the introducing notations are used below
\begin{eqnarray}
I=\int\,\varepsilon_{abcd}\,\left\{-\frac{1}{l^2_P}R^{ab}\wedge e^c\wedge e^d+
\frac{1}{6}\,\Theta^a\wedge e^b\ \wedge e^c\wedge e^d\,
\right\}\,, \label{30}
\end{eqnarray}
\begin{eqnarray}
\d\omega^{ab}+\omega^a_c\wedge\omega^{cb}=\frac{1}{2}\,R^{ab}\,,
\nonumber \\
\Theta^a=\frac{i}{2}\,\big(\,\opsi\,\gamma^a\, {\cal
D}_{\mu}\psi-{\cal D}_{\mu}\opsi\,\gamma^a\,\psi\,\big)\,\d
x^{\mu}\,.
\label{40}
\end{eqnarray}
Here,
\begin{eqnarray}
\omega^{ab}=\omega^{ab}_{\mu}\,\d x^{\mu}
\label{50}
\end{eqnarray}
is a connection 1-form in a certain orthonormal basis
$\{e^{\mu}_a\}$, \ $e^a=e^a_{\mu}\,\d x^{\mu}$, \
$g_{\mu\nu}=e_{\mu}^a\,e_{\nu}^a$ is a metric tensor, $l_P$ is the
Planck length, $\psi$ is a Dirac field, $\gamma^a$ are the Dirac
matrices, and ${\cal D}_{\mu}\psi$ is the covariant derivative of
the Dirac field.

The most characteristic property of the $B$-$F$-theory consists in that
the curvature tensor 2-form is equal to zero. However, while the $B$-$F$-theory really
describes gravity in a three-dimensional space, this is not so in
higher dimensional spaces. For example, in a four-dimensional
space with the action (\ref{30}), the curvature tensor is not equal to zero.
But it is not the only difficulty of the theory: the introduction of the matter
is also awkward.

The detailed description of discrete quantum theory of gravity
based on the $B$-$F$ formalism can be found in \cite{5}-\cite{14}.

In contrast to the multidimensional case, considerable
computational progress has been made in two-dimensional discrete
quantum gravity (see \cite{15}, \cite{16}). We refer the reader to
\cite{17} for the relation between the three-dimensional quantum
Yang-Mills theory on a lattice and three-dimensional gravity.

In the present paper, we propose a new version of discrete quantum
theory of gravity. This new theory differs both from the Regge
theory and from the discrete variant of the $B$-$F$-theory. Just
as in the $B$-$F$-theory, the connection in our theory is
represented by the elements a gauge group. However, unlike the
$B$-$F$-theory, all fundamental variables in the theory proposed
here are defined directly on the elements of the simplicial
complex itself. In particular, the gauge group $G$ elements that
play the role of connection 1-forms are defined on one-dimensional
simplices. In contrast to the $B$-$F$-theory, we explicitly
introduce an analogue of a tetrad 1-form in our theory. The
elements of a tetrad 1-form are also defined on one-dimensional
simplices and belong to the real vector space with the basis
formed by Dirac matrices. The presence of a tetrad form in the
theory allows us to introduce a Dirac field whose elements are
defined at the vertices of a simplicial complex and they are
transformed by a spinor representation of the group $G$. Using
these fields, one can easily construct a lattice action which is
invariant relative to the gauge transformations. It is important
that this action obviously manifests the continuum limit, at least
at "naive" level. In the naive continuum limit, this action is
reduced to a standard action of the continuum theory of gravity
(\ref{30}). Further the quantization of discrete gravity is
performed. This means the determination of the fundamental
partition function, which represents a functional gauge-invariant
integral over the introduced fields. It is shown qualitatively
that this quantum theory displays the tendency to degenerate into
macroscopic continuum theory. It turns out that the correct
determination of the partition function requires that the gauge
group should be compact, which is equivalent to a metric with
Euclidean signature. Euclidean signature in fundamental quantum
gravity has been introduced much earlier in the work \cite{18}.

It seems very interesting that in highly compressed state of
universe (which is possible at small times near the singularity) a
sort of "high-temperature expansion" is valid. Therefore, the
confinement of "color" takes place at early stage of universe
expansion. Note that in the considered theory any nontrivial
representation of the local Lorentz or gauge group $G$ (i.e.
spinor, vector and so on fields) plays the role of color.

In conclusion, the arguments are given in favour of the dynamics
of discrete quantum gravity leads to the significant loosening of
packing of field modes in momentum space. Thus, the packing of
modes in momentum space in the corresponding continuum theory turns
out to be significantly noncompact. From here it follows that the
universe has the weight smaller by many orders in comparison with
the case of the usual quantum field theory.

Here we use the results obtained in the previous publications
\cite{19}, \cite{20}.

It is necessary mention here two interesting works \cite{21}, \cite{22},
in which the formulations of discrete quantum gravity are the most close
to the one presented here. But the constructions in \cite{21}, \cite{22}
are based on the hypercubical lattices but not the simplicial complexes.

\section{Discrete Quantum Gravity. Definition of Action}

Let $\mK$ be a 4-dimensional simplicial complex admitting
geometrical realization. The definition and required properties of
simplicial complexes can be found in \cite{19}. A detailed theory
of simplicial complexes is given, for example, in
\cite{23} -- \cite{24}. Below instead of "simplicial complex" we say simply
"complex", and the concepts in the following pairs are treated as
synonyms: 0-simplex and vertex; 1-simplex and edge; 2-simplex and
triangle; 3-simplex and tetrahedron. The finite complexes with a
4-disk topology are interesting here. Such complexes have a
boundary $\partial\mK$ which is 3-dimensional complex with
topology of 3-sphere $S^3$. Denote by $\alpha_q,\,\, q=0,1,2,3,4$
the number of q-simplexes of the complex $\mK$. The indexes
$i,j,k,l,\,\ldots$ run through the complex vertices: $a_i,\,\,a_j$
and so on. Two vertices are called adjacent if these two vertices
are the boundary vertices of the same edge.

For convenience I give here the definition of orientation of simplexes and
complexes.

A simplex
\begin{eqnarray}
s^r=\varepsilon(a_0,\,a_1,\,\ldots\,,\,a_r)\equiv
\varepsilon\,a_0\,a_1\,\ldots\,a_r
\label{discr3}
\end{eqnarray}
has an orientation, or is oriented, if every order of its vertices is
assigned a sign "+" or "-", so that orders differing by an odd permutation correspond to opposite signs.
Thus if $\varepsilon=1$ the orientation of simplex (\ref{discr3}) is given by the
orders $(a_0,\,a_1,\,\ldots\,,\,a_r)$ or $-(a_1,\,a_0,\,\ldots\,,\,a_r)$.
Let $(a_0,\,\ldots\,,\,a_{i-1},\,a_{i+1},\,\ldots\,,\,a_r)$ be the face of a simplex
$s^r$ obtained by eliminating one vertex $a_i$ from the sequence
$a_0,\,a_1,\,\ldots\,,\,a_r$. By definition, the orientation of this face, given by
\begin{eqnarray}
B^{r-1}_i=(-1)^i\,\varepsilon(a_0,\,
\ldots\,,\,a_{i-1},\,a_{i+1},\,\ldots\,,\,a_r)\,,
\label{discr4}
\end{eqnarray}
is called an iduced orientation of the simplex $s^r$.

Denote by $D$ the maximum value of number $r$ in
(\ref{discr3}) for all simplexes of complex. In the
considered case $D=4$. Thus $D$ is the dimension of
complex. Two oriented $D$-dimensional simplices $s^D_1$ and
$s^D_2$ of a $D$-dimensional simplicial complex are called
concordantly oriented if either the simplices $s^D_1$ and
$s^D_2$ have no common $(D-1)$-dimensional faces or the
orientation of their common $(D-1)$-dimensional face
$B^{D-1}$ induced by the orientation of the simplex $s^D_1$
is opposite to the orientation of the same face $B^{D-1}$
induced by the orientation of the simplex $s^D_2$. A
$D$-dimensional simplicial complex $\mK$ is called
orientable if there exists such an orientation for all its
$D$-dimensional simplices that any pair of its
$D$-dimensional simplices is concordantly oriented. The
concordant orientation of $D$-dimensional simplices defines
the orientation of the complex, and namely this orientation
of $D$-simplices is regarded as positive.

Evidently, interesting for us the complex $\mK$ is
orientable.

Below index $A$ enumerates 4-simplices. Introduce the
following notation for oriented 1-simplices in the case
when the vertexes $a_i$ and $a_j$ belong to the 4-simplex
with index $A$
\begin{eqnarray}
X^A_{ij}=a_ia_j=-X^A_{ji}\,.
\label{discr5}
\end{eqnarray}

Let
\begin{eqnarray}
s^4_A=a_{i_0}a_{i_1}a_{i_2}a_{i_3}a_{i_4}
\label{discr6}
\end{eqnarray}
be an positively oriented 4-simlex with index $A$.
An oriented frame of a simplex (\ref{discr6}) at a vertex $a_{i_0}$ is
the ordered set of four oriented 1-simplices (\ref{discr5}) such that an
even permutation of these 1-simplices does not change the orientation
while an odd permutation changes the orientation of the frame to the opposite.
By definition, the frame
\begin{eqnarray}
{\cR}^{A\,i_0}=\big(X^A_{i_0i_1},\,X^A_{i_0i_2},\,X^A_{i_0i_3},\,X^A_{i_0i_4}\big)
\label{discr7}
\end{eqnarray}
is oriented positively.

Let $\gamma^a,\,\,a,b,c,\ldots =1,2,3,4$ be $4\times 4$ Dirac
matrices with Euclidean signature. Thus all Dirac matrices as well as
matrix
\begin{eqnarray}
\gamma^5=\gamma^1\gamma^2\gamma^3\gamma^4\,, \qquad
\tr\,\gamma^5\gamma^a\gamma^b\gamma^c\gamma^d=4\,\varepsilon
^{abcd}
\label{discr10}
\end{eqnarray}
are Hermitian. To each vertex $a_i$, we assign the Dirac spinors
$\psi_i$ and $\overline{\psi}_i$ each of whose components assumes
values in a complex Grassman algebra. In the case of Euclidean
signature, the spinors $\psi_i$ and $\overline{\psi}_i$ are
independent variables and are interchanged under the Hermitian
conjugation. The Dirac matrixes act from the left to the spinors
$\psi_i$ and from the right to the spinors $\overline{\psi}_i$.

Let us assign to each oriented edge $a_ia_j$ an element of the
group  $Spin(4)$
\begin{eqnarray}
\Omega_{ij}=\Omega^{-1}_{ji}=\exp\left(\frac{1}{2}\omega^{ab}_{ij}
\sigma^{ab}\right)\,, \ \ \ \sigma^{ab}=\frac{1}{4}[\gamma^a,\,\gamma^b]\,.
\label{discr20}
\end{eqnarray}
Holonomy element $\Omega_{ij}$ of the gravitational field executes
a parallel transformation of spinor $\psi_j$ from vertex $a_j$ of
edge $a_ia_j$ to neighboring vertex $a_i$. We denote by $V$ a
linear space with basis $\gamma^a$. Let each oriented edge
$a_ia_j$ be put in correspondence with element $\hat{e}_{ij}\equiv
e^a_{ij}\gamma^a\in V$, such that
\begin{eqnarray}
\hat{e}_{ij}=-\Omega_{ij}\hat{e}_{ji}\Omega_{ij}^{-1}\,.
\label{discr30}
\end{eqnarray}
The notations
$\overline{\psi}_{Ai}, \ \psi_{Ai}, \ \hat{e}_{Aij}, \
\Omega_{Aij}$ and so on indicate that edge $X^A_{ij}=a_ia_j$ belongs to 4-simplex
with index $A$. Here, the sign "$ \ - \ $" in (\ref{discr30}) is
due to the fact that $e_{A\,ij}$ and $e_{A\,ji}$ are the values of the 1-form on the
edges $X^A_{ij}=a_ia_j$ and $X^A_{ji}=a_ja_i=-a_ia_j=-X^A_{ij}$ (which are oriented mutually oppositely), respectively.
The "facing" from the elements of a holonomy group on the right-hand side of
Eq. (\ref{discr30}) are necessary since the element
$e_{A\,ji}$ must be paralled-translated from the vertex $a_{A\,j}$ to the vertex
 $a_{A\,i}$ to compare this element with the element $e_{A\,ij}$. The quantities
 assigned to each oriented edge $a_ia_j$ and satisfying to Eq. (\ref{discr30})
 are called 1-forms.

By assumption, the complex $\mK$ has a disk topology. For such a
complex, the concept of orientation can be introduced. We define
the orientation of the complex by defining the orientation of each
4-simplex. In this case, if two 4-simplices have a common
tetrahedron, the two orientations of the tetrahedron, which are
defined by the orientations of these two 4-simplices, are
opposite. In our case, the complex obviously has only two
orientations.

Let $a_{Ai}, \ a_{Aj}, \ a_{Ak}, \ a_{Al}$, and $a_{Am}$ be all
five vertices of a 4-simplex with index $A$ and
$\varepsilon_{Aijklm}=\pm 1$ depending on whether the order of
vertices $a_{Ai}\,a_{Aj}\,a_{Ak}\,a_{Al}\,a_{Am}$ defines the
positive or negative orientation of this 4-simplex. In addition,
$\varepsilon_{ijklm}=0$ if at least two indices coincide. We can
now write the Euclidean action in the model in question
\begin{eqnarray}
I=\frac{1}{5\times
24}\sum_A\sum_{i,j,k,l,m}\varepsilon_{Aijklm}\tr\,\gamma^5 \times
\nonumber \\
\times\left\{-\frac{1}{2\,l^2_P}\Omega_{Ami}\Omega_{Aij}\Omega_{Ajm}
\hat{e}_{Amk}\hat{e}_{Aml}+\right.
\nonumber \\
\left.+\frac{1}{24}\hat{\Theta}_{Ami}
\hat{e}_{Amj}\hat{e}_{Amk}\hat{e}_{Aml}\right\}\,,
\label{discr40}
\end{eqnarray}
\begin{gather}
\hat{\Theta}_{Aij}=
\frac{i}{2}\gamma^a\left(\overline{\psi}_{Ai}\gamma^a
\Omega_{Aij}\psi_{Aj}-\overline{\psi}_{Aj}\Omega_{Aji}\gamma^a\psi_{Ai}\right)
\equiv
\nonumber \\
\equiv\Theta^a_{Aij}\gamma^a\in V\,.
\label{discr50}
\end{gather}
The quantity $\hat{\Theta}_{Aij}$, as well as the whole
action (\ref{discr40}), represents an Hermitean operator.
One can easily verify that 1-form (\ref{discr50}), just as
the 1-form $\hat{e}_{ij}$, satisfies relation
(\ref{discr30}). This fact is established by the repeated
application of the formula
\begin{gather}
S^{-1}\,\gamma^a\,S=S^a_b\,\gamma^b\,,
\label{discr60}
\end{gather}
where
\begin{gather}
S\equiv\exp\frac{1}{2}\,\varepsilon_{ab}\,\sigma^{ab}\,, \qquad \
\varepsilon_{ab}=-\varepsilon_{ba}=\varepsilon^a_b\,,
\nonumber \\
S^a_b\equiv\big(\exp\varepsilon\big)^a_b=
\delta^a_b+\varepsilon^a_b+\frac{1}{2}\,
\varepsilon^a_c\varepsilon^c_b+\,\ldots\,.
\label{discr70}
\end{gather}

The volume of a 4-complex is given by
\begin{gather}
V_A=\frac{1}{4!}\times\frac{1}{5!}\times
\nonumber \\
\times\sum_A\sum_{i,j,k,l,m}\varepsilon_{A\,ijklm}\,
\varepsilon^{abcd}\,e^a_{A\,mi}e^b_{A\,mj}e^c_{A\,mk}e^d_{A\,ml}\,.
\label{discr80}
\end{gather}
Here, factor $1/4!$ is required since the volume of a
four-dimensional parallelepiped with generatrices
$e^a_{A\,mi},\;e^b_{A\,mj},\;e^c_{A\,mk}$, and $e^d_{A\,ml}$ is
$4!$ times larger than the volume of a 4-simplex with the same
generatrices, while factor $1/5!$ is due to the fact that all five
vertices of each simplex are taken into account independently.

The dynamic variables are quantities $\Omega_{ij}$ and
$\hat{e}_{ij}$, which describe the gravitational degrees of
freedom, and fields $\overline{\psi}_i$ and $\psi_i$, which are
material fermion fields (other material fields are not considered
here).

In the space of fields, there acts a gauge group according to the
following rule. To each vertex $a_{A\,i}$, let us assign an element of
the group $S_{A\,i}\in\Spin(4)$. According to the principle of gauge
invariance, the fields $\Omega$, \ $e$, \
$\psi$, and the transformed fields
\begin{gather}
\tilde{\Omega}_{A\,ij}=S_{A\,i}\,\Omega_{A\,ij}\,S^{-1}_{A\,j}\,,
\nonumber \\
\tilde{e}_{A\,ij}=S_{A\,i}\,e_{A\,ij}\,S^{-1}_{A\,i}\,,
\nonumber \\
\tilde{\psi}_{A\,i}=S_{A\,i}\,\psi_{A\,i}\,, \ \qquad \
\tilde{\overline{\psi}}_{A\,i}=\overline{\psi}_{A\,i}\,S^{-1}_{A\,i}
\label{discr90}
\end{gather}
are physucally equivalent. This means that the action (\ref{discr40}) is invariant
under the transformations (\ref{discr90}). Under the gauge transformatios
(\ref{discr90}), the 1-form $\Theta$ is transformed in the same way as the form
$e$
\begin{gather}
\tilde{\hat{\Theta}}_{A\,ij}=S_{A\,i}\,\hat{\Theta}_{A\,ij}\,S^{-1}_{A\,i}\,.
\label{discr100}
\end{gather}
The last formula is verified with the help of Eqs.
(\ref{discr60}), (\ref{discr70}) and (\ref{discr90}). Gauge
invariance of the action (\ref{discr40}) and of the volume
(\ref{discr80}) is established by using Eqs.
(\ref{discr90}) and (\ref{discr100}).

It is natural to interpret the quantity
\begin{gather}
l^2_{ij}\equiv\frac{1}{4}\,\tr\,(\hat{e}_{ij})^2=\sum_{a=1}^4(e^a_{ij})^2
\label{discr110}
\end{gather}
as the square of the length of the edge $a_ia_j$. Thus, the geometric properties
of a simplicial complex prove to be completely defined.

Now, let us show that, in the limit of slowly varying fields, the action
(\ref{discr40}) reduces to the continuum action of gravity, minimally
connected with with a Dirac field, in a four-dimensional Euclidean space.

Consider a certain subset of vertices from the simplicial complex and assign
the coordinates (real numbers)
\begin{gather}
x^{\mu}_{A\,i}\equiv x^{\mu}(a_{A\,i})\,,
 \qquad \ \mu=1,\,2,\,3,\,4
\label{discr120}
\end{gather}
to each vertex $a_{A\,i}$ from this subset. We stress that these coordinates
are defined only by their vertices rather than by the higher dimension simplices
to whom these vertices belong; moreover, the correspondence between the vertices
from the subset considered and the coordinates (\ref{discr120}) is one-to-one.

Suppose that
\begin{gather}
|\,x^{\mu}_{A\,i}-x^{\mu}_{A\,j}\,|\sim a\,,
\label{discr130}
\end{gather}
where the parameter $a$ is of the order of the lattice spacing.
Estimates (\ref{discr130}) has the sense if the
complex contains a very large number of simplices and its
geometric realization is an almost smooth four-dimensional
surface {\footnote{Here, by an almost smooth surface, we
mean a piecewise smooth surface consisting of flat
four-dimensional simplices, such that the angles between
adjacent 4-simplices tend to zero and the sizes of these
simplices are commensurable.}}. Suppose also that the four
4-vectors
\begin{gather}
\d x^{\mu}_{A\,ji}\equiv x^{\mu}_{A\,i}-x^{\mu}_{A\,j}\,,
 \ \ \qquad i\neq j\,, \ \ \ i=1,\,2,\,3,\,4
\label{discr140}
\end{gather}
are linearly independent and
\begin{gather}
\left\vert
\begin{array}{llll}
\d x^1_{A\,m1} \ & \ \d x^2_{A\,m1} \ & \ldots & \ \d x^4_{A\,m1}\\
\ldots  & \ldots  &  \ldots  & \ldots \\
\d x^1_{A\,m4} \ & \ \d x^2_{A\,m4} \ & \ldots & \ \d x^4_{A\,m4}
\end{array}\right\vert >0\,,
\label{discr150}
\end{gather}
provided that the frame $\big(X^A_{m\,1},\,
\ldots\,,\,X^A_{m\,4}\,\big)$ is positively oriented. Inequality
(\ref{discr150}) implies that positively oriented local coordinates
are introduced on the almost flat surface considered. Here, the differentials of coordinates
(\ref{discr140}) correspond to one-dimensional simplices $a_{A\,j}a_{A\,i}$, so that,
if the vertex $a_{A\,j}$ has coordinates $x^{\mu}_{A\,j}$, then the vertex
$a_{A\,i}$ has the coordinates $x^{\mu}_{A\,j}+\d x^{\mu}_{A\,ji}$.

In the continuum limit, the holonomy group elements (\ref{discr20}) are
close to the identity element, so that the quantities $\omega^{ab}_{ij}$
tend to zero being of the order of $O(\d x^{\mu})$.
Thus one can consider the following system of equation for $\omega_{A\,m\mu}$
\begin{gather}
\omega_{A\,m\mu}\,\d x^{\mu}_{A\,mi}=\omega_{A\,mi}\,,  \ \ \
i=1,\,2,\,3,\,4\,.
\label{discr160}
\end{gather}
In this system of linear equation, the indices $A$ and $m$ are
fixed, the summation is carried out over the index $\mu$, and
index runs over all its values. Since the determinant
(\ref{discr150}) is positive, the quantities $\omega_{A\,m\mu}$
are defined uniquely. Suppose that a one-dimensional simplex
$X^A_{m\,i}$ belong to four-dimensional simplices with indices
$A_1,\,A_2,\,\ldots\,,\,A_r$. Introduce the quantity
\begin{gather}
\omega_{\mu}\left[\frac{1}{2}\,(x_{A\,m}+
x_{A\,i})\,\right]\equiv\frac{1}{r}\,
\bigg\{\omega_{A_1\,m\mu}+\,\ldots\,+\omega_{A_r\,m\mu}\,\bigg\}\,,
\label{discr170}
\end{gather}
which is assumed to be related to the midpoint of the segment
$[x^{\mu}_{A\,m},\,x^{\mu}_{A\,i}\,]$. Recall that the coordinates
$x^{\mu}_{A\,i}$ just as the differentials (\ref{discr140}),
depend only on vertices but not on the higher dimensional
simplices to which these vertices belong. According to the
definition, we have the following chain of equalities
\begin{gather}
\omega_{A_1\,mi}=\omega_{A_2\,mi}= \,\ldots\,=\omega_{A_r\,mi}\,.
\label{discr180}
\end{gather}
It follows from (\ref{discr140}) and
(\ref{discr160})--(\ref{discr180}) that
\begin{gather}
\omega_{\mu}\left(x_{A\,m}+ \frac{1}{2}\,\d x_{A\,mi}\,\right)\,\d
x^{\mu}_{A\,mi}=\omega_{A\,mi}  \,.
\label{discr190}
\end{gather}
The value of the field $\omega_{\mu}$ in (\ref{discr190}) on each
one-dimensional simplex is uniquely defined by this simplex.

Next, we assume that the fields $\omega_{\mu}$ smoothly depend on
the points belonging to the geometric realization of each
four-dimensional simplex. In this case, the following formula is
valid up to $O\big((\d x)^2\big)$ inclusive
\begin{gather}
\Omega_{A\,mi}\,\Omega_{A\,ij}\,\Omega_{A\,jm}=
\exp\left[\frac{1}{2}\,\mR_{A\,m\mu\nu}\,\d x^{\mu}_{A\,mi}\, \d
x^{\nu}_{A\,mj}\,\right]\,,
\label{discr200}
\end{gather}
where
\begin{gather}
\mR_{A\,m\mu\nu}=\partial_{\mu}\omega_{A\,m\nu}-\partial_{\nu}\omega_{A\,m\mu}+
[\omega_{A\,m\mu},\,\omega_{A\,m\nu}\,]\,.
\label{discr210}
\end{gather}
On the right-hand side of (\ref{discr200}), as well as in equality
(\ref{discr210}), all fields are taken at the vertex $a_{A\,m}$ of
a four-dimensional simplex $A$ as is indicated by the subscript
$A\,m$. When deriving formula (\ref{discr200}), we used the
Hausdorff formula.

In exact analogy with (\ref{discr160}), let us write out the following relations
for a tetrad field without explanations
\begin{gather}
e_{A\,m\mu}\,\d x^{\mu}_{A\,mi}=e_{A\,mi}\,.
\label{discr220}
\end{gather}

Using (\ref{discr20}) and (\ref{discr160}), we can rewrite the 1-form
(\ref{discr50}) as
\begin{gather}
\Theta_{A\,ij}=\gamma^a\,\frac{i}{2}\,\left[\,
\overline{\psi}_{A\,i}\,\gamma^a\,{\cal D}_{\mu}\,\psi_{A\,i}-
\overline{{\cal D}_{\mu}\,\psi}_{A\,i}\,\gamma^a\,\psi_{A\,i}\,
\right]\,\d x^{\mu}_{A\,ij}\,,
\label{discr230}
\end{gather}
to within $O(\d x)$; here,
\begin{gather}
{\cal D}_{\mu}\,\psi_{A\,i}=\partial_{\mu}\,
\psi_{A\,i}+\omega_{A\,i\mu}\,\psi_{A\,i}\,.
\label{discr240}
\end{gather}

Before rewriting the action (\ref{discr40}) in the continuum limit,
we give the following obvious formula
\begin{gather}
\sum_{\sigma(Am)}\,\varepsilon_{\sigma(Am)}\,
\d x^{\mu}_{A\,mi}\,\d x^{\nu}_{A\,mj}\,\d x^{\lambda}_{A\,mk}\,
\d x^{\rho}_{A\,ml}=
\nonumber \\
=24\,\varepsilon^{\mu\nu\lambda\rho}\,
v_{S\,A}\,.
\label{discr250}
\end{gather}
Here, $\varepsilon^{\mu\nu\lambda\rho}$ is a completely antisymmetric symbol,
which is equal to unity when $(\mu\,\nu\,\lambda\,\rho)=(1\,2\,3\,4)$ (compare
with (\ref{discr150})), and $v_{S\,A}$ is the volume of the geometric realization
of simplex $A$ in a four-dimensional Euclidean space when the Euclidean coordinates
of the geometric realization of the simplex are equal to the corresponding
coordinates of its vertices (\ref{discr120}). The factor 24 in (\ref{discr250})
is necessary since the volume $v_{S\,A}$ of the four-dimensional simplex
on the right-hand side is less than the volume of a four-dimensional parallelepiped
constructed on the vectors $\d x^{\mu}_{A\,mi},\,\ldots\,,\,\d x^{\mu}_{A\,ml}$
by a factor of 24.

Applying formulas (\ref{discr200})--(\ref{discr250}),  changing the
summation to integration and taking into account that
\begin{gather}
\mR\equiv\frac{1}{2}\,\sigma^{ab}\,
R^{ab}_{\mu\nu}\,\d x^{\mu}\wedge\d x^{\nu}\,, \quad
e=\gamma^a\,e^a_{\mu}\,\d x^{\mu}\,,
\nonumber \\
\Theta=\gamma^a\,\frac{i}{2}\left[\,\overline{\psi}
\gamma^a\,{\cal D}_{\mu}\,\psi-\overline{{\cal D}_{\mu}\psi}
\,\gamma^a\,\psi\,\right]\,\d x^{\mu}\,,
\label{discr270}
\end{gather}
we obtain the well known expression (\ref{30}) for the
action (\ref{discr40}) in the continuum limit:

Thus, in the naive continuum limit, the action
(\ref{discr40}) proves to be equal to the action of gravity
with a $\Lambda$-term and a metric with Euclidean signature
that is minimally coupled to a Dirac field.

\section{Quantization of Discrete Gravity}

Let us determine the partition function $Z$ for a discrete
Euclidean gravity \footnote{We mean here that the partition
function is a functional of the values of the dynamic
variables at the boundary $\partial\mK$.}. Let us enumerate
the zero-dimensional (vertices) and one-dimensional (edges)
simplices by indices $\cV$ and $\cE$, respectively, and
denote by $\psi_{\cV}$, \ $\Omega_{\cE}$, etc. the
corresponding variables. By definition,
\begin{gather}
Z=\const\cdot\bigg (\prod_{\cE}\,\int\, \d\Omega_{\cE}\,\int\,\d
e_{\cE}\,\bigg)\times
\nonumber \\
\times\big(\prod_{\cV}\,\d\overline{\psi}_{\cV}\,
\d\psi_{\cV}\,\big)\,\exp\big(-I\,\big)\,. \label{discr280}
\end{gather}
Here, $\const$ is a certain normalizing factor, $\d\Omega_{\cE}$
is the Haar measure on the group $\Spin(4)$,
\begin{gather}
\d e_{\cE}\equiv\prod_a\,\d e^a_{\cE}\,, \qquad \
e_{\cE}=e^a_{\cE}\,\gamma^a\,, \label{discr290}
\end{gather}
and
\begin{gather}
\d\overline{\psi}_{\cV}\,\d\psi_{\cV}\equiv\prod_{\nu}\,
\d\overline{\psi}_{\cV\nu}\,\d\psi_{\cV\nu}\,.
\label{discr300}
\end{gather}
The index $\nu$ in (\ref{discr300}) enumerates individual components of the
spinors $\psi_{\cV}$ and $\overline{\psi}_{\cV}$, such that we have
a product of the differentials of all independent generators of the
Grassman algebra of Dirac spinors in (\ref{discr300}). The action $I$ in
(\ref{discr280}) is defined by formula
(\ref{discr40}).

Note that the measure (\ref{discr290}) is determined correctly in view
of invariance of the Haar measure and the relations
(\ref{discr20}) and (\ref{discr30}). Therefore, one can really assume
that the measure (\ref{discr290}) is related to the set of edges.

Obviously, all the measures used in the functional integral
(\ref{discr280}) are invariant under the gauge transformations (\ref{discr90}).
Since the action $I$ (\ref{discr40}) in (\ref{discr280}) is also
gauge invariant, the partition function (\ref{discr280}) is invariant
under the action of the gauge group (\ref{discr90}).

Consider the partition function (\ref{discr280}) with a zero
$\Lambda$-term in the absence of fermions. In this case, the
integral over the 1-form $e_{\cE}$ becomes Gaussian
\begin{gather}
Y\big\{\Omega\,\big\}=\int\,D\,z\cdot\exp
\left(\frac{1}{2}\,z_m\,{\ccm}_{m\,n}\,z_n\right)\,.
\label{discr310}
\end{gather}
Here, $\{\,z_m\,\}$, \ $m=1,\,\ldots\,,\,Q$ denotes a set of real
variables $\{\omega^a_{\cE}\}$ and ${\ccm}_{mn}$ is a real symmetrical
matrix depending on the elements of the holonomy group $\Omega_{\cE}$.
Thus,
\begin{gather}
\frac{1}{2}\,z_m\,{\ccm}_{mn}\,z_n\equiv
\frac{1}{l^2_P}\,\frac{1}{5}\cdot\frac{1}{24}\,\sum_{A,\,m}\,
\sum_{\sigma(A\,m)}\,\varepsilon_{\sigma(Am)}\times
\nonumber \\
\times\tr\big(\gamma^5\,\Omega_{A\,mi}\,
\Omega_{A\,ij}\,\Omega_{A\,jm}\,e_{A\,mk}\,e_{A\,ml}\,\big)\,.
\label{discr320}
\end{gather}

Denote by $\{\,\lambda_q\,\}$, where $q=1,\,\ldots\,,\,Q$,
a set of eigenvalues of the matrix ${\ccm}_{mn}$. Let
$\varepsilon_q=\sign\lambda_q$. Since, in general, there are both
negative and positive eigenvalues among $\{\lambda_q\}$, the integral
(\ref{discr310}) should be redefined. This is done by passing to Lorentzian
signature. Under this procedure, the eigenvalues are transformed by
the rule
$$
\lambda_q\rightarrow e^{i\varphi}\,\lambda_q\,,
$$
where $\varphi=0$ in the Euclidean space and $\varphi=\pi/2$
in the case of the Minkowski signature. Thus, the Euclidean
Gaussian integral
\begin{gather}
{\cI}_E=\frac{1}{\sqrt{2\pi}}\,\int_{-\infty}^{+\infty}\,
\d z\cdot\exp\left(\frac{1}{2}\,\lambda\,z^2\,\right)
\label{discr330}
\end{gather}
reduces to the Fresnel integral in the Minkowski signature
\begin{gather}
{\cI}_M=\frac{1}{\sqrt{2\pi}}\,\int_{-\infty}^{+\infty}\,
\d z\cdot\exp\left(\frac{i}{2}\,\lambda\,z^2\,\right)=
\sqrt{\frac{i}{\lambda}}=
\nonumber \\
=(i)^{\frac{\varepsilon}{2}}\,
\frac{1}{\sqrt{|\,\lambda\,|}} \,,
\label{discr340}
\end{gather}
where $\varepsilon=\sign\lambda$. Let us perform the analytic continuation
$$
\lambda\rightarrow
e^{-i\varphi}\,\lambda
$$
on the right-hand side of Eq. (\ref{discr340}) and set $\varphi=\pi/2$.
Thus, we recover the Euclidean signature of a metric and
obtain the following value for integral (\ref{discr330})
\begin{gather}
{\cI}_E=(i)^{\frac{\varepsilon+1}{2}}\,
\frac{1}{\sqrt{|\,\lambda\,|}}\,.
\label{discr350}
\end{gather}

Now, using Eq. (\ref{discr350}), we redefine the integral
(\ref{discr310}) of interest
\begin{gather}
Y\big\{\Omega\,\big\}=\const\,\prod_q\,
(i)^{\frac{\varepsilon_q+1}{2}}\,|\,\lambda_q\,|^{-1/2}\,.
\label{discr360}
\end{gather}

If there are fermion fields in the theory, one should first
calculate a functional integral over fermions. The subsequent
integration over the 1-form $e$ remains Gaussian and yields a
contribution of the form (\ref{discr360}) to the partition
function. The remaining integral over the elements of the holonomy
group $\Omega$ may prove to be divergent despite the compactness
of this group. Indeed, certain eigenvalues $\lambda_q$ may vanish
under certain configurations of the field $\Omega$. Since the
expression under the integral sign depends on the negative powers
of $\lambda_q$, the integral over the field $\Omega$ may prove to
be divergent. From the physical point of view, these divergences
are of great interest. Note that the tendency of eigenvalues
$\lambda_q$ to zero implies that the integral over the 1-form
$e^a$ is saturated when the absolute values of this field $e^a$
(or its certain components) tend to infinity. This means that the
size of universe tends to infinity (see (\ref{discr110})). On the
other hand, as will be shown below, the fact that the field
components $e^a$ have large values implies that the dynamics of
the system becomes quasiclassical. Therefore, from the physical
viewpoint, these divergences imply birth of quasiclassical
macroscopic space-time.

Concerning the problem under discussion, we note that the presence
of Dirac fields in integral (\ref{discr280}) only strengthens the
divergence under the integration over the field $e^a$. Indeed,
after the integration over the fermion field, the integral over
the field $e^a$ is rewritten as (cf. (\ref{discr330}) and
(\ref{discr340}))
\begin{gather}
{\cI}=\frac{1}{\sqrt{2\pi}}\,\int_{-\infty}^{+\infty}\, \d
z\,P_n(z)\cdot\exp\left(\frac{i}{2}\,\lambda z^2\,\right)\,,
 \label{discr370}
\end{gather}
where $P_n(z)$ is a polynomial in $z$ of degree $n$. For small
$\lambda$, integral (\ref{discr370}) is proportional to
$|\,\lambda\,|^{-(n+1)/2}$.

A similar physical interpretation of divergences under the
integration over the field $e^a$ in the continuum quantum
$B$-$F$-theory in a three-dimensional space-time was given by
Witten in \cite{25}.

Let us notice another possible type of divergences in a discrete
quantum gravity. If the partition function (\ref{discr280}) was
defined for a metric with Lorentzian signature, then the elements
of the holonomy group would be the noncompact group
$\Spin(3,\,1)$. The gauge group (\ref{discr90}) would also be
noncompact, being a direct product of the $\cV$ copies of the
group $\Spin(3,\,1)$. Since both the measure and action in the
transfer-matrix are gauge invariant, the functional integral in
the transfer-matrix would not be defined at all before the
fixation (at least partial) of the gauge. However, the fixation of
the gauge in the fundamental transfer-matrix seems to be a so
artificial procedure that the theory itself looses its beauty and
sense. In our opinion, this means that the fundamental partition
function for a discrete theory of gravity can be constructed only
on the basis of a metric with Euclidean signature.

In their well-known paper \cite{18}, Hartle and Hawking made a
hypothesis that the wave function of the universe must be
calculated with the use of the functional integral on the basis of
a metric with Euclidean signature. But in the case of the gravity
theory the Euclidean action is not positively defined. In our
opinion, the arguments for a metric with Euclidean signature
provided by the discrete theory of gravity are much more reliable
than the arguments given in \cite{18}.

\section{High Temperature Expansion}

From the beginning let us consider the integral (\ref{discr280}) in the
region of integration variables where
\begin{gather}
|e_{ij}^a|>l_0\gg l_P\,. \label{discr380}
\end{gather}
In this region each item in the sum (\ref{discr40}) generally is
also large since the items in the sum (\ref{discr40}) are
polynomials in the variables $e_{ij}^a$ of powers not less than
two. Therefore the whole integral in (\ref{discr280}) can be
estimated quasiclassically or by the stationary phase method. In
this region one must use the long wave limit action (\ref{30}),
and to perform the stationary phase calculations the integration
paths in (\ref{discr280}) must be deformed so that Lorentzian
signature is realized. We see that in the
considered model the time arises dynamically in continuum limit.
The study of continuum limit of the theory is performed in the
subsequent sections.

Now let us consider the integral (\ref{discr280}) in the
region of integration variables where
\begin{gather}
|e_{ij}^a|<l_1\ll l_P\,.
\label{discr390}
\end{gather}
In this region each item in the sum (\ref{discr40}) is small, so that
the subintegral quantity in (\ref{discr280}) (in the case of pure gravity
and zero $\Lambda$-term) can be written as
\begin{gather}
\exp\big(-I\,\big)=\prod_{A}\prod_{i,j,k,l,m}
\bigg(1+\frac{1}{5\times24\times2\times
l^2_P}\,\varepsilon_{Aijklm}\tr\,\gamma^5 \times
\nonumber \\
\times\Omega_{Ami}\Omega_{Aij}\Omega_{Ajm}
\hat{e}_{Amk}\hat{e}_{Aml}+\ldots\bigg)\,. \label{discr400}
\end{gather}
The expansion (\ref{discr400}) is called here as
high-temperature expansion. It is well known that the
analogous representation for the $\exp\big(-I\,\big)$ is
true in the lattice Yang--Mills theory in the limit of
large coupling constant. From such representation the
significant phenomenon of colour confinement follows.
Originally the phenomenon of colour confinement has been
obtained analytically with the help of high-temperature
expansion (with the help of representation of the tipe
(\ref{discr400})) by Wilson, and then numerous computer
simulations confirmed this conclusion. Since the situations
concerning high-temperature expansion in both theories are
closely analogous, we make the conclusion that in the
region of variables (\ref{discr390}) also take place colour
confinement. Introduce the following notations
$$
C=\{a_{i_0}a_{i_1},\,a_{i_1}a_{i_2},\ldots\,,a_{i_r}a_{i_0}\}
$$
is a closed contour or a one dimensional subcomplex with zero boundary;
$$
W\,(C)=\langle\,\tr\left(\Omega_{i_0i_1}\Omega_{i_1i_2}\ldots \Omega_{i_ri_0}\right)\,\rangle_1
$$
is Wilson loop correlator which in our case is calculated in
the theory of pure gravity with zero $\Lambda$-term in the region
of variables restricted by inequalities (\ref{discr390}). Let $\sigma_C$ be a
two dimensional subcomplex with boundary $\partial\,\sigma=C$ and
$n_C(\sigma)$ be the number of triangles containing in $\sigma_C$, and
$$
n_C=\min_{\sigma}\{n_C(\sigma)\}\,.
$$
Then the simple standard calculations give the following estimation
\begin{gather}
W\,(C)\sim\exp\left(-n_C\mu\,\ln l_1^{-1}\right)\,.
\label{discr410}
\end{gather}
Here $\mu$ is a positive number which does not depend on contour $C$ and parameter $l_1$.

Let us emphasize that in the case of discrete quantum gravity the role
of colour gauge group plays the group (\ref{discr90}). Thus only singlet
(i.e. scalar, but not spinor, vector and so on) fields
with respect to the group (\ref{discr90}) have quasiparticle excitations
in the region (\ref{discr390}), i.e. on the early stages of universe development.
This conclusion partially justifies the use only scalar fields in
numerous works in which the dynamics of early universe is investigated.
But in contrast to the Yang--Mills theory in expanding universe
 the phase transition occurs to
deconfinement phase (formally in the region (\ref{discr380})). In
this phase the dynamics becomes quasiclassical.

Another variant of
high temperature expansion, but with the same confined quantum numbers,
is presented in \cite{21}.

\section{The qualitative behaviour of elliptic operators spectrum
on random breathing lattice}

Let us now show that the modes of quantized fields in the
quasiclassical continuum phase have essentially noncompact packing
in momentum space. This important conclusion follows from
high-temperature expansion and the most general properties of
spectrum of elliptic operators.

We illustrate the effect in Appendix A on the example of the
spectrum of one dimensional discrete Laplace operator on random
lattice on a cycle. In the cases of 3 and 4 vertexes the problem
is solved exactly and we see that in the case when the total
length of the cycle is fixed but the distances between some
vertexes tend to zero some of eigenvalues of the operator tend to
infinity as an inverse first or second power of the small
distances between the corresponding vertexes. This phenomenon of
eigenvalues noncompact packing in the case of strongly random
lattice I call "spectrum loosening".

Now let us consider the eigenfunction problem for Dirac operator
on three-dimensional complex $\mS=\partial\mK$ and observe the
spectrum loosening as a consequence of randomness of the lattice.
In our case, let index $A$ enumerates tetrahedra. Analogously to
the four-dimensional case, $\varepsilon_{A\,ijkl}=\pm 1$ depending
on whether the order of vertices $a_{Ai}a_{Aj}a_{Ak}a_{Al}$
defines the positive or negative orientation of the corresponding
tetrahedron.

In the three-dimensional case the fermion part of the action (cf.
formula (\ref{discr40})) can be written as
\begin{gather}
I\{\opsi,\,\psi,\,e\}\equiv I_{\psi}=\frac{1}{2\cdot
4\cdot6}\sum_A\sum_{i,j,k,l}\varepsilon_{A\,lijk}
\varepsilon^{abc}\,\Theta^a_{A\,li}e^b_{A\,lj}e^c_{A\,lk}\,.
\label{discr412}
\end{gather}
Since the volume of three-dimensional complex (compare with Eq.
(\ref{discr80}))
\begin{gather}
V=\frac{1}{3!}\cdot\frac{1}{4!}\sum_A\sum_{i,j,k,l}\varepsilon_{A\,lijk}
\varepsilon^{abc}e^a_{Ali}e^b_{Alj}e^c_{Alk}\,, \label{discr413}
\end{gather}
the scalar product in the space of Dirac field modes
\begin{gather}
\langle\psi_P|\psi_Q\rangle=\frac{1}{3!}\cdot\frac{1}{4!}\sum_A\sum_{i,j,k,l}
\opsi_{P\,Al}\psi_{Q\,Al}\,\varepsilon_{A\,lijk}
\varepsilon^{abc}e^a_{Ali}e^b_{Alj}e^c_{Alk}\,. \label{discr414}
\end{gather}
Here the indices $P$ and $Q$ enumerate the modes of Dirac field.
Evidently, in continuum limit
\begin{gather}
\langle\psi_P|\psi_Q\rangle\longrightarrow\int\d^3x\,\opsi_P(x)\psi_Q(x)\,.
\label{discr415}
\end{gather}

Let us consider the problem of extremum of quadratic form
(\ref{discr412}) under the condition $\langle\psi|\psi\rangle=1$.
According to the Lagrange method this problem is equivalent to the
following one
\begin{gather}
\frac{\delta}{\delta\opsi_i}\bigg(I\,\{\opsi,\,\psi,\,e\}-\epsilon\langle\psi|\psi\rangle\bigg)=0\,.
\label{discr416}
\end{gather}
Let $\{\psi_{P\,i}\}$ be a complete orthonormal set of solutions
of Eq. (\ref{discr416}), so that
\begin{gather}
\langle\psi_P|\psi_Q\rangle=\delta_{P\,Q}\,, \qquad
I\{\opsi_P,\,\psi_P,\,e\}=\epsilon_P\,.
\label{discr417}
\end{gather}
Using Eqs. (\ref{discr416}) and (\ref{discr417}), for any
configuration of field
\begin{gather}
\psi_i=\sum_Pc_P\psi_{P\,i}\,,
\nonumber
\end{gather}
we find:
\begin{gather}
\frac{I\{\opsi,\,\psi,\,e\}}{\langle\psi|\psi\rangle}=\frac{\sum_P|c_P|^2\,\epsilon_P}{\sum_P|c_P|^2}\,.
\nonumber
\end{gather}
From here the following evident estimation follows
\begin{gather}
\left|\frac{I\{\opsi,\,\psi,\,e\}}{\langle\psi|\psi\rangle}\right|\leq
|\epsilon_P|_{max}\,.
\label{discr418}
\end{gather}
In the right hand side of the last inequality the quantity
$|\epsilon_P|_{max}$ designates the maximum value among the set of
values $\{|\epsilon_P|\}$.

Although complex $\mS$ has the topology of 3-sphere $S^3$, this
circumstance does not play significant role in the present
consideration. Therefore, taking into account that
$\alpha_q\rightarrow\infty$, we shall consider any small part of
complex $\mS$ as a three-dimensional complex $\mS'$ embedded in a
three-dimensional Euclidean space. To simplify the consideration,
we will assume that
\begin{gather}
\Omega_{ij}=1\,\,, \ \ \
\bigl(e_{ij}^a+e_{jk}^a+\ldots+e_{li}^a\bigr)=0\,\,.
\label{discr411}
\end{gather}
Here, the sum in the parentheses is taken over any closed path
formed by 1-simplices belonging to complex $\mS'$ and
$\Omega_{ij}$ is the connection assigned to the 1-simplex
$a_ia_j\in\mS'$. Equations (\ref{discr411}) indicates that the
curvature and torsion are equal to zero. Thus, the complex $\mS'$
is in the three-dimensional Euclidean space, $e^a_{ij}$ being the
components of the vector in a certain orthogonal basis in this
space, and if $R^a_i$ is the radius-vector of vertex $a_i$, then
$e_{ij}^a=R^a_j-R^a_i$.

Let's show that some values of $|\epsilon_P|$ increase without
limit on breathing lattice with fixed total volume of subcomplex
$\mS'$. Under the "breathing lattice" I understand here the
situation when the quantities $\{e_{\cE}\}$ are independent
dynamical variables. This means, of course, that the wave function
of universe depends also on the set of variables $\{e_{\cE}\}$,
and averaging-out over quantum fluctuations includes also
averaging-out over the introduced higher values $\{R^a_i\}$

To advance further, let's perform some consideration.

Write out the part of fermion action depending only on spinors
associated with neighboring vertexes $a_i$ and $a_j$ belonging to
the boundary of 1-simplex $a_ia_j$. Let $s^3_A, \, A=1,\,\ldots
\,,\,n$ be the complete set of tetrahedrons, each of which
contains 1-simplex $a_{Ai}a_{Aj}$. Denote by
\begin{gather}
\mv=\cup_{A=1,\ldots,\,n}s^3_A\in\mS
\nonumber
\end{gather}
the minimal three-dimensional subcomplex such that
$a_ia_j\notin\partial\mv$, but $a_i\in\partial\mv$ and
$a_j\in\partial\mv$. Let $a_k\in\partial\mv, \, k=1,\,\ldots
\,,\,n$ be the rest of vertexes of complex $\mv$ which does not
coincide with $a_i$ and $a_j$ and $a_ka_{k+1}$ be a 1-simplex
belonging to $\partial\mv$. Thus it is clear that 1-simplex
$a_na_1$ exists and it belong to the boundary $\partial\mv$. For
convenience, we assume that index $k$ is defined in $(\!\!\!\!\mod
n$), so that $a_na_{n+1}=a_na_1$. It is assumed also that the
motion
\begin{gather}
a_1\rightarrow a_2\rightarrow\,\ldots\,\rightarrow a_n\rightarrow
a_1
\nonumber
\end{gather}
appears to be positive (counterclockwise) to an "observer" located
at vertex $a_j$. Introduce the notation
\begin{gather}
S^a_{ij}=\frac12\sum_{k=1}^n\varepsilon^{abc}e^b_{jk}e^c_{j,k+1}\,.
\nonumber
\end{gather}
It is not difficult to show that the part of fermion action
depending only on spinors associated with neighboring vertexes
$a_i$ and $a_j$ is
\begin{gather}
I_{\psi\,ij}=\frac{i}{12}\overline{\psi}_i
\gamma^aS^a_{ij}\psi_j+h.c.\,.
\label{discr419}
\end{gather}

Let the configuration $\{\psi_k\}$ be such that $\psi_k=0$ for
$k\neq i$ and $k\neq j$, and
\begin{gather}
\opsi_i\psi_i\sim\opsi_j\psi_j\sim 1\,.
\label{discr421}
\end{gather}
In other words, the spinors $\psi_k$ differ from zero only on
vertexes $a_i$ and $a_j$. We assume also that in the vicinity of
vertexes $a_i$ and $a_j$
\begin{gather}
l^2_{kl}<a^2
\label{discr422}
\end{gather}
(see formula (\ref{discr110})). Then, combining the formulae
(\ref{discr414}), (\ref{discr418}) and (\ref{discr419}), we find
the following estimation
\begin{gather}
\left|\frac{I\{\opsi,\,\psi,\,e\}}{\langle\psi|\psi\rangle}\right|\sim\frac{1}{a}\,.
\label{discr423}
\end{gather}
Thus we see that
\begin{gather}
|\epsilon_P|_{max}\sim
a^{-1}\;\;\;{}_{\overrightarrow{a\rightarrow 0}} \;\;\;\infty\,.
\label{discr424}
\end{gather}
One must emphasize that the total volume  (\ref{discr413}) of the
space $\mS$ is maintained constant when the parameter $a$ in
(\ref{discr422}) and (\ref{discr424}) tends to zero.

Let's pay attention to the well known result that on periodic lattice with the
lattice spacing $a$ the estimation (\ref{discr424}) also is valid.
But in this case the limit in (\ref{discr424}) is impossible if
the total volume and the number of vertexes of the lattice are
fixed.

Now one needs take into account the fact that the lattice
in our case is not only random but it is also breathing
since the quantities $e^a_{ij}$ are the dynamical
variables. Further we keep in mind the scalar field since
the spinor structure does not affect significantly the
estimation.

Firstly, we write out the trivial formula for the volume in
momentum space occupied by $N$ modes placed in the flat volume $V$
and densely packed in momentum space
\begin{gather}
\Omega=(2\pi)^3\frac{N}{V}\,.
\label{discr420}
\end{gather}
One can easily come to Eq. (\ref{discr420}) by analysing the
discrete Laplace equation on periodic lattice. At first let's
consider for simplicity one-dimensional lattice, the vertexes of
which are numerated in series by the whole numbers
$n=0,\,1,\,\ldots\,,\,N\gg1$. The complete set of orthonormal
eigenfunctions of discrete Laplace operator we take as
$$
\psi_k(n)=\frac{1}{\sqrt{N}}e^{ikn}, \qquad \psi_k(0)=\psi_k(N).
$$
From here we obtain the quantization conditions for the
quasi-momenta
$$
k_l=\frac{2\pi l}{N}, \qquad l=0,\,\pm1,\,\ldots\,,\,\pm(N/2).
$$
Thus, the maximal difference of the quasi-momentum values is
$$
\Delta_{max}k=\frac{2\pi}{N}\Delta_{max}l=2\pi.
$$
Now let's pass to the dimensional quantities by introducing
the lattice spacing $a$. Then $L=Na$ is the length or
volume of the space, and $\Delta_{max}p=\Delta_{max}k/a$ is
the volume in the momentum space which is occupied by the
whole set of modes. Thus we have
\begin{gather}
\Delta_{max}p\equiv\Omega=\frac{\Delta_{max}k}{a}=2\pi\frac{N}{L}.
\label{discr421}
\end{gather}
Obviously, the quantity (\ref{discr421}) is the minimal volume in
momentum space which can be occupied by the complete set of
momenta of $N$ independent modes of the Laplace operator in
concidered problem. Equation (\ref{discr420}) is obtained now by
exponentiation of the right hand side of Eq. (\ref{discr421}) in
third power.

Take into account the fact that in confinement phase all
correlators of fundamental fields drop exponentially with space
separation. Moreover, the correlators of color fields in
space-time representation are proportional to $\delta$-functions.
This means that the fields at nearest regions of space volume are
not correlated. It is natural to assume that the same conclusion
remains true at initial times in quasiclassical phase. Therefore
let us divide a macroscopic volume $V$ with the total number of
degrees of freedom (or the number of modes) $N$ into ${\cal N}$
subvolumes $v_i$ in each of which $n_i$ degrees of freedom is
contained. Thus
\begin{gather}
\sum_{i=1}^{{\cal N}}n_i=N\,, \qquad  \sum_{i=1}^{{\cal N}}v_i=V\,,
\label{discr430}
\end{gather}
and
\begin{gather}
\omega_i=(2\pi)^3\frac{n_i}{v_i}
\label{discr440}
\end{gather}
is the minimal possible volume in momentum
space occupied by $n_i$ modes placed in the flat volume $v_i$.
Now instead of the quantity (\ref{discr420}) one must consider the
following quantity
\begin{gather}
\tilde{\Omega}=\frac{(2\pi)^3}{{\cal N}}\sum_{i=1}^{{\cal N}}\frac{n_i}{v_i}\,.
\label{discr450}
\end{gather}
Indeed, the minimum of quantity (\ref{discr450}) subjected to the
constraints (\ref{discr430}) is equal to (\ref{discr420}).

Since in the considered theory the volumes $v_i$ are variable
quantities, one must introduce the measure on the manifold of
volumes $\{v_i\}$. The simplest measure agreeing with fundamental
measure (\ref{discr290}) looks like as follows
\begin{gather}
\d\mu=\frac{({\cal N}-1)!}{V^{{\cal N}-1}}\delta\left(V-\sum_{i=1}^{{\cal N}}v_i\right)
\prod_{i=1}^{{\cal N}}\d v_i\,, \quad v_i>0\,,
\nonumber \\
\int\d\mu=1\,.
\label{discr460}
\end{gather}
To justify the measure (\ref{discr460}) we give the following
arguments:

1) The elementary volumes are given by Eq. (\ref{discr80}) from
which it is seen that the volumes are determined only by 1-forms
$e^a_{ij}$.

2) The variables $\{e^a_{ij}\}$ change independently in integral
(\ref{discr280}).

3) It is important here that the action (\ref{discr40}) remains
almost unchanged under the changes of the variables $\{e^a_{ij}\}$
on a large range. This assertion becomes rigorous in continuum
limit if only smooth long-waves are taken into account with
wavelengths mich greather than lattice spacing. Indeed, the
1-forms $\omega_{Am\mu}$ and $e_{Am\mu}$ in Eqs. (\ref{discr160})
and (\ref{discr220}) do not change under the change of the
right-hand sides of these equations and simultaneous changes of
the differentials $\d x^{\mu}_{Ami}$. In other words, the quantity
$e^a_{ij}$ is the value of differential form $e^a_{\mu}\d x^{\mu}$
on the vector $e^a_{ij}$. But the continuum action (\ref{30})
depends on the fields $e^a_{\mu}, \, \omega^a_{\mu}$. Thus we see
that the changes of the elementary volumes in Eq. (\ref{discr460})
due to the changes of the variables $\{e^a_{ij}\}$ weakly affect
the continuum action. Hence, the wave function remains almost
unchanged under the changes of the variables $\{e^a_{ij}\}$ on a
large range. Let's designate by
\begin{gather}
{\cal D}e\equiv\prod_{{\cE}\in\mS}\prod_a\,\d e^a_{\cE}
\label{discr461}
\end{gather}
the functional measure with the help of which the scalar products
of wave functions are calculated. The measure (\ref{discr461}) can
be factorized
\begin{gather}
{\cal D}e={\cal D}e_1\d\mu.
\label{discr462}
\end{gather}
By definition, the submeasure ${\cal D}e_1$ in (\ref{discr462})
does not depend on the volume variables $\{v_i\}$ and, conversely,
the quantity (\ref{discr450}) does not depend on the variables
determining the submeasure ${\cal D}e_1$. \\\\

Hence, instead of (\ref{discr450}) the more physically sensible
quantity is
\begin{gather}
\langle\tilde{\Omega}\,\rangle\equiv\int\tilde{\Omega}\d\mu=(2\pi)^3\frac{{\cal
N}-1}{V\,{\cal N}}\sum_{i=1}^{{\cal N}}n_i\int_{v_i\ll V}\frac{\d
v_i}{v_i}=
\nonumber \\
=(2\pi)^3\frac{N}{V}\int_{v_i\ll V}\frac{\d v_i}{v_i}\,.
 \label{discr470}
\end{gather}
The last equality is obtained at taking into account the first
constraint of (\ref{discr430}) and the relation ${\cal N}\gg 1$.

The comparison of Eqs. (\ref{discr420}) and (\ref{discr470}) shows
that taking into account the dynamics of the system leads to the
essential expansion of the momentum space volume occupied by
quantum field modes. This expansion factor is
\begin{gather}
\varkappa_1=\int_{v_i\ll V}\frac{\d
v_i}{v_i}=3\ln\frac{a_1}{a_0}=3\ln\xi_0\,.
\label{discr480}
\end{gather}
Here $a_0$ is some minimal dimension of the theory and $a_1\ll
V^{1/3}$. It seems that $a_0\gg l_P$, since only at $|e_{ij}^a|\gg l_P$
the quasiclassical phase can exist (see (\ref{discr380})).

But it is not the end of story. From the obtained
estimation (\ref{discr480}) we elaborate a kind of
renormalization group describing loosening of mode packing.
Let $n$ be the number of steps of renormalization group and
\begin{gather}
\xi_s=\frac{a_{s+1}}{a_s}=\xi\gg 1\,, \quad s=1,\ldots\,,n\,,
\label{discr490}
\end{gather}
and $a_{n+1}=a$ is the radius of universe. Thus
$\xi^n=\xi_1\xi_2\ldots\xi_n=a/a_0$. For rough estimation let us
take
\begin{gather}
n=\frac{1}{\lambda}\ln\frac{a}{a_0}\gg 1\,, \quad \lambda\gg 1\,.
\label{discr500}
\end{gather}
Using Eqs. (\ref{discr480})--(\ref{discr500}) it is easy to see
that the expansion factor of momentum space volume occupied by
modes after $n$ steps is
\begin{gather}
\varkappa_n=\prod_{s=1}^n(3\ln\xi_s)=(3\ln\xi)^n=\left(\frac{a}{a_0}\right)^{(\ln
3\lambda)/\lambda}\,.
\label{discr510}
\end{gather}
The value of right hand side of Eq. (\ref{discr510}) can bee very
large (many orders) in magnitude. This phenomenon is called here
as "spectrum loosening". It seems that the effect of spectrum
loosening and translational invariance are compatible only on
breathing lattice.

The more careful study of the problem is possible only if
we take into account the dynamics of the system.

\section{Conclusion}

It follows from the presented analysis, that the continuum quantum
gravity arising from the discrete quantum gravity (if it exists)
possess very unusual properties. For example, let's try to estimate the
contribution to cosmological constant due to quantum field fluctuations
in the framework of presented here theory. We shall see that the unsolvable
problem of a large value of cosmological constant can be solved if the estimation
(\ref{discr510}) is taken into account.

It is well known the estimation for the cosmological constant in the
framework of the usual quantum field theory (see, for example \cite{26})
\begin{gather}
\lambda_{eff}\sim G\,\Lambda^4\sim l_P^2\Lambda^4\,.
\label{discr520}
\end{gather}
Here $G$ is the Newtonian gravitational constant, $l_P$ is the Planck space-time
scale, and $\Lambda$ is the cutoff parameter in momentum space which
usually is restricted by the Planck scale: $\Lambda\sim l_P^{-1}$. Thus the
estimation (\ref{discr520}) becomes as follows
\begin{gather}
\lambda_{eff}\sim l_P^{-2}\,.
\label{discr530}
\end{gather}
It should be noted that this estimation is valid i) independently on the size
of universe and ii) under the tacit assumption of compact packing of the field modes
in momentum space.

But in the elaborated here theory one should correct the estimation (\ref{discr530})
by the factor $\varkappa_n^{-1}\lll 1$ (see (\ref{discr510})) for the reason of noncompact
packing of the field modes in momentum space! Thus, instead of estimation
(\ref{discr530}) now we have the following one
\begin{gather}
\lambda_{eff}\sim \left(\frac{a_0}{a}\right)^{(\ln
3\lambda)/\lambda}l_P^{-2}\lll l_P^{-2}\,.
\label{discr540}
\end{gather}

So the effective cosmological constant can be made enough small.
Thus a possibility of solving the cosmological constant problem arises.
This possibility will be elaborated in the subsequent papers.

\begin{acknowledgments}

This work was
supported by the Program for Support of Leading Scientific Schools
$\sharp$ 2044.2003.2. and RFBR $\sharp$ 04-02-16970-a.

\end{acknowledgments}

\appendix

\section{}

Let us consider the discrete Laplace operator on a one dimensional
cycle with 3 vertexes (see fig. 1). The numbers $a$, $b$, $c$
are the distances between the vertexes 1 and 2, 2 and 3, 3 and 1,
correspondingly. In the vertexes 1, 2 and 3 the real numbers
$\varphi_1$, $\varphi_2$ and $\varphi_3$ are defined. Write out the discrete
equation for Laplace operator eigenfunctions
\begin{gather}
-(\Delta\varphi)_1=-\frac{2}{ac}\left(\frac{a\,\varphi_3+c\,\varphi_2}{a+c}
-\varphi_1\right)=\epsilon\,\varphi_1\,,
\nonumber \\
-(\Delta\varphi)_2=-\frac{2}{ab}\left(\frac{b\,\varphi_1+a\,\varphi_3}{a+b}
-\varphi_2\right)=\epsilon\,\varphi_2\,,
\nonumber \\
-(\Delta\varphi)_3=-\frac{2}{bc}\left(\frac{c\,\varphi_2+b\,\varphi_1}{b+c}
-\varphi_3\right)=\epsilon\,\varphi_3\,.
\label{ap1}
\end{gather}
For enough smoothly varying (from vertex to vertex) variables $\varphi_i$
the system of equations (\ref{ap1}) transforms to the continuum equation
$-\Delta\,\varphi=\epsilon\,\varphi$.
The three eigenvalues of Eq. (\ref{ap1}) are as follows
\begin{gather}
\epsilon_1=0\,,
\nonumber \\
\epsilon_{2,3}=\frac{a+b+c}{abc}\left[1\pm\sqrt{1-\frac{8\,abc}{(a+b)(a+c)(b+c)}}\,\right]\,.
\label{ap2}
\end{gather}
If $a\rightarrow 0$ and $(a+b+c)=\const$, then
\begin{gather}
\epsilon_2\sim\frac{2(b+c)}{abc}\longrightarrow\infty\,, \qquad
\epsilon_3=\frac{4}{bc}\,.
\label{ap3}
\end{gather}
{\psfrag{Ph1}{\kern-5pt\lower-1pt\hbox{\large $\phi_1$}}
\psfrag{Ph2}{\kern0pt\lower0pt\hbox{\large $\phi_2$}}
\psfrag{Ph3}{\kern0pt\lower0pt\hbox{\large $\phi_3$}}
\psfrag{Ia}{\kern0pt\lower0.5pt\hbox{\large $a$}}
\psfrag{Ib}{\kern0pt\lower0.5pt\hbox{\large $b$}}
\psfrag{Ic}{\kern0pt\lower0.5pt\hbox{\large $c$}}
\begin{figure}[tbp]
 \includegraphics[width=0.20\textwidth]{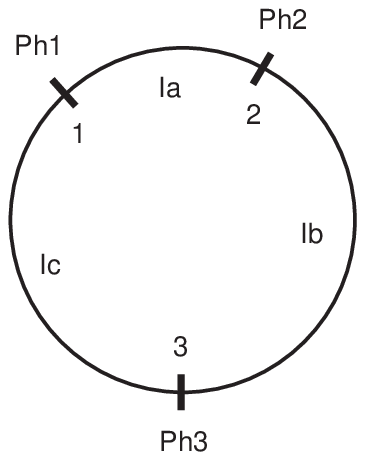}
\caption{}
\label{One}
 \end{figure}}


Consider the same problem for the discrete Laplace operator on a one dimensional
cycle with 4 vertexes separated in order by distances
$a$, $b$, $c$ and $d$. Then the eigenvalues of the operator satisfy
the following equation
\begin{gather}
\epsilon^4-2\epsilon^3\left(\frac{1}{cd}+\frac{1}{bc}+\frac{1}{ab}+\frac{1}{ad}\right)+
\nonumber \\
+4\epsilon^2\left[\frac{1}{bc^2d}+\frac{1}{ab^2c}+\frac{1}{acd^2}+\frac{1}{a^2bd}+\frac{2}{abcd}-\right.
\nonumber \\
\left.-\frac{1}{c^2(b+c)(c+d)}-\frac{1}{b^2(a+b)(b+c)}-\right.
\nonumber \\
\left.-\frac{1}{a^2(a+b)(a+d)}
-\frac{1}{d^2(a+d)(c+d)}\right]-
\nonumber \\
-8\epsilon\left[\frac{1}{ab^2c^2d}+\frac{1}{abc^2d^2}+\frac{1}{a^2bcd^2}+\frac{1}{a^2b^2cd}-\right.
\nonumber \\
\left.-\frac{a+c}{ab^2cd(a+b)(b+c)}-\frac{b+d}{a^2bcd(a+d)(a+b)}-\right.
\nonumber \\
\left.-\frac{b+d}{abc^2d(b+c)(c+d)}-\frac{a+c}{abcd^2(a+d)(c+d)}\right]=0\,.
\label{ap4}
\end{gather}
Though the exact solution we did not obtain,
the approximate solutions of this equation in two interesting here
special cases we write out
\begin{gather}
b=d=l\,, \qquad a\rightarrow 0\,, \qquad c\rightarrow 0\,:
\nonumber \\
\epsilon_1=0\,, \quad \epsilon_2\approx\frac{4}{l^2}\,,
\nonumber \\
\epsilon_3\approx\frac{4}{la}\rightarrow\infty\,, \quad
 \epsilon_4\approx\frac{4}{lc}\rightarrow\infty\,.
\label{ap5}
\end{gather}
\begin{gather}
c=d=l\,, \qquad  a\rightarrow 0\,, \qquad b\rightarrow 0\,:
\nonumber \\
\epsilon_1=0\,, \quad \epsilon_{2,3}\approx\frac{2}{l(a+b)}\rightarrow\infty\,,
\nonumber \\
\epsilon_4\approx\frac{2}{ab}-\frac{4}{l(a+b)}\rightarrow\infty\,.
\label{ap6}
\end{gather}

\end{document}